\documentclass[superscriptaddress,showpacs,preprintnumbers,amsmath,amssymb,prl,twocolumn]{revtex4-1}
\usepackage{graphicx}% Include figure files
\usepackage{dcolumn}% Align table columns on decimal point
\usepackage{bm}% bold math
\usepackage{color}
\usepackage{amsmath}
\usepackage{amssymb}
\usepackage{enumerate}
\usepackage{epstopdf}
\usepackage{comment}
\usepackage{keyval}%
\usepackage{fancyhdr}

\graphicspath{{./Figures/}}
\def\nba#1{}               % accepted \nba so it stays in the source but disappears on compiling

\begin{document}

%\preprint{APS/123-QED}

%================================================================================================
\title{Evidence for Fluctuating Charge Stripes far above the Charge-Ordering Transition  in La$_{1.67}$Sr$_{0.33}$NiO$_{4}$ }
%================================================================================================
\author{A. M. Milinda Abeykoon}
\affiliation{Condensed Matter Physics and Materials Science Department, Brookhaven National Laboratory, Upton, New York 11973, USA\\}

\author{Emil~S. Bo\v{z}in}
\affiliation{Condensed Matter Physics and Materials Science Department, Brookhaven National Laboratory, Upton, New York 11973, USA\\}

\author{Wei-Guo Yin}
%\email{wyin@bnl.gov}
\affiliation{Condensed Matter Physics and Materials Science Department, Brookhaven National Laboratory, Upton, New York 11973, USA\\}

\author{Genda Gu}
\affiliation{Condensed Matter Physics and Materials Science Department, Brookhaven National Laboratory, Upton, New York 11973, USA\\}

\author{John P. Hill }
\affiliation{Condensed Matter Physics and Materials Science Department, Brookhaven National Laboratory, Upton, New York 11973, USA\\}

\author{John M. Tranquada}
\affiliation{Condensed Matter Physics and Materials Science Department, Brookhaven National Laboratory, Upton, New York 11973, USA\\}

\author{Simon J. L. Billinge}
\affiliation{Condensed Matter Physics and Materials Science Department, Brookhaven National Laboratory, Upton, New York 11973, USA\\}
\affiliation{Department of Applied Physics and Applied Mathematics, Columbia University, New York, NY 10027, USA}

%------------------------------------------------------------------------------------------------------------------------

\begin{abstract}

\noindent
The temperature evolution of structural effects associated with charge (CO) and spin order (SO) in La$_{1.67}$Sr$_{0.33}$NiO$_{4}$  has been investigated using neutron powder diffraction. We report an anomalous shrinking of the $c/a$ lattice parameter ratio that correlates with $T_{\rm CO}$. The sign of this change can be explained by the change in interlayer Coulomb energy between the static stripe-ordered state and the fluctuating stripe-ordered state or the charge-disordered state.  In addition, we identify a contribution to the mean-square displacements of Ni and in-plane O atoms whose width correlates quite well with the size of the pseudogap extracted from the reported optical conductivity, with a non-Debye-like component that persists below and well above $T_{\rm CO}$.  We infer that dynamic charge-stripe correlations survive to $T\sim2T_{\rm CO}$

\end{abstract}

%------------------------------------------------------------------------------------------------------------------------

\date{\today}% It is always \today, today,


\maketitle

%------------------------------------------------------------------------------------------------------------------------

The phenomenology of charge and spin stripes has been of considerable interest with regard to the problem of high-temperature superconductivity in cuprates \cite{kive03,vojt09,zaan01}.  Although static charge-stripe order has been detected by diffraction techniques only in select cuprate materials \cite{huck12,abba12,fuji12b}, stripe-like modulations of electronic states have been detected at the surface of cleavable cuprates by scanning tunneling spectroscopy (STS) \cite{robe06,kohs07,park10}, and a new type of charge correlation has recently been identified in YBa$_2$Cu$_3$O$_{6+x}$ \cite{ghir12,chan12a,blac13,blan13,2013arXiv1305.5515T}.  When static charge-stripe order occurs in La$_{2-x}$Ba$_x$CuO$_4$ \cite{huck11}, it competes with Josephson coupling between the superconducting layers \cite{taji01,li07,berg09b,home12}.  Dynamic stripes may have a broader relevance to superconductivity in the cuprates \cite{kive98}; however, there is not yet any direct, incontrovertible evidence for dynamic charge stripes in cuprates.

La$_{2-x}$Sr$_x$NiO$_4$ has served as a model system for studying stripe order \cite{ulbr12b}.  Charge-stripe order can occur at relatively high temperatures, reaching a maximum of $T_{\rm CO} = 240$~K at $x=\frac13$ \cite{rami96,lee97}.  While the stripe order is more classical than in the cuprates, with quantum fluctuations being unimportant, there is the possibility that dynamical stripes might exist above $T_{\rm CO}$.  Inelastic neutron scattering studies have already shown that spin-stripe correlations survive above both $T_{\rm SO}$ and $T_{\rm CO}$ \cite{lee02}.  An x-ray scattering study was able to follow critical scattering associated with charge-stripe correlations to temperatures slightly above $T_{\rm CO}$, but ran out of signal by $T_{\rm CO}+20$~K \cite{du;prl00}.  In contrast, measurements of optical conductivity have demonstrated the absence of a dominant Drude peak, the signature of a conventional metallic state, to temperatures as high as $2T_{\rm CO}$ \cite{katsu;prb96}.  The fact that the optical conductivity remains peaked at finite frequency suggests that the charge carriers are quasi-localized, which might be the result of fluctuating charge stripes persisting at high temperature.

In this paper, we use neutron scattering from a polycrystalline sample to investigate the structural response associated with charge-stripe correlations in La$_{1.67}$Sr$_{0.33}$NiO$_{4}$ (LSNO).  This might appear to be a surprising approach, as even in the stripe-ordered state, the superlattice peaks associated with stripe order are too weak to detect by powder diffraction.  Nevertheless, we observe distinct signatures in the temperature-dependent scattering associated with the loss of static charge stripe order at $T_{\rm CO}$ and the gradual attenuation of fluctuating stripes at much higher temperatures.

%------------------------------------------------------------------------------------------------------------------------

The La$_{1.67}$Sr$_{0.33}$NiO$_{4}$ sample studied here was initially prepared as a single crystal by the traveling-solvent floating-zone method; a portion of the crystal was then ground to powder form. Neutron diffraction measurements on a piece of the starting crystal have confirmed that $T_{\rm CO}=240$~K, consistent with previous work \cite{lee97,rami96,du;prl00}. Time-of-flight neutron powder diffraction measurements were carried out at the NPDF beamline~\cite{proff;apa01i} of the Los Alamos Neutron Scattering Center (LANSCE) at Los Alamos National Laboratory (LANL). Data were converted to the pair distribution function (PDF) using PDFgetN~\cite{peter;jac00} and the PDF was modeled using PDFgui~\cite{farro;jpcm07}.   Rietveld refinements were performed using GSAS/EXPGUI~\cite{larso;unpub04,toby;aca04}. The tetragonal space group, $I4/mmm$, was used~in the refinements \cite{wu02}. \nba{reference with the structure model, seminal reference if possible} More details of the experiments and data reduction may be found in the supplementary materials.

\begin{figure}[t]%[tb]
\center
\includegraphics[scale=0.55,angle=0]{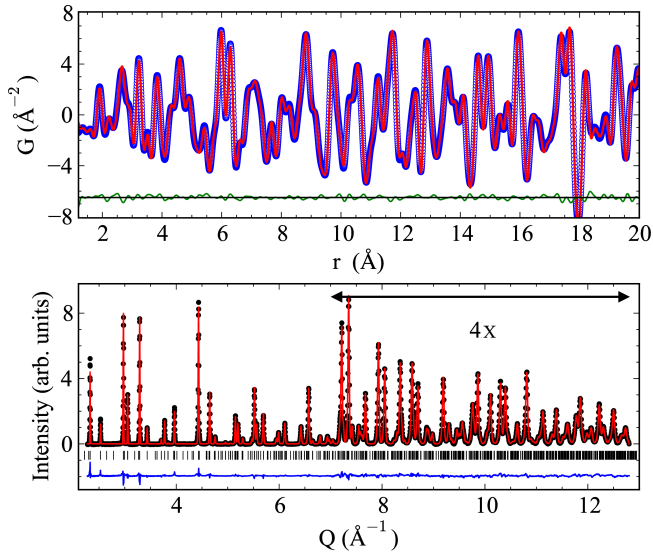}
\caption {(Color online) PDF (top) and Rietveld (bottom) fits to 80~K data. In both panels, structure models are shown as red solid lines, while the data are shown as circles. Difference curves are offset for clarity.}
\label{fig:PDFRit_fits}
\end{figure}

\begin{figure}[t]
\center
\includegraphics[scale=0.73]{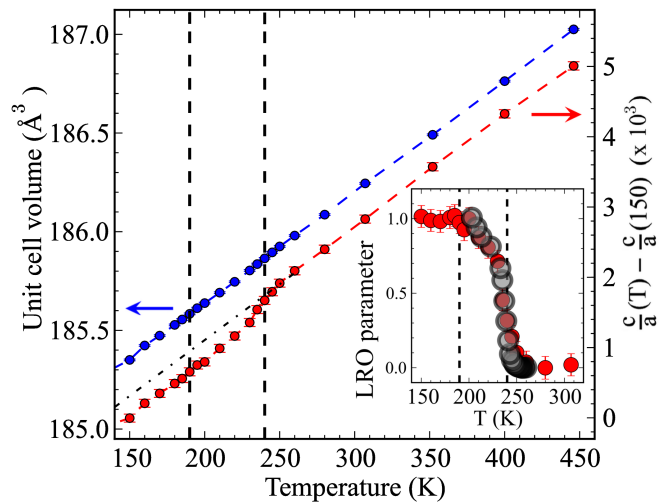}
\caption {(Color online) The temperature evolution of $c/a$ ratio (red solid  circles) in comparison with the volume expansion (blue solid circles). Dotted lines indicate $T_{\rm SO}=190$~K and $T_{\rm CO}=240$~K. In the inset, the red solid circles represent a normalized order parameter calculated from $c/a$, and the black open circles represents the $T$ evolution of the normalized  integrated intensity of the single crystal CO peak from Du et.al.~\cite{du;prl00}.}
\label{fig:LSNO_ca}
\end{figure}

Representative Rietveld and PDF fits to the 80~K data are shown in Figure~\ref{fig:PDFRit_fits}.
The model fit to the data is the undistorted tetragonal structure, which does not allow for the average atomic displacements associated with charge ordering.  The fits are good, which indicates that any structural modification due to the charge ordering is small, consistent with earlier studies \cite{wu02}.  No charge or spin superlattice peaks are evident in the data; even in the best case of a nickelate with 3D long-range order, the superlattice intensities are no more than $10^{-4}$ relative to strong fundamental reflections \cite{tran95b}, which is beyond the dynamic range for powder diffraction.
%
%======================
%======================
%
While we do not detect the average atomic displacements associated with charge ordering, the transition is revealed by a modification of the tetragonality of the unit cell.
The unit cell volume varies smoothly with temperature, as shown in Fig.~\ref{fig:LSNO_ca} by the blue circles, in agreement with earlier studies~\cite{Guoqi;prb02};
however, the $c$/$a$ ratio  exhibits a strong anomaly at $T_{\rm CO}$. This anomaly  results from a contraction in the $c$  and an elongation in the $a$ directions in a way that preserves the total unit cell volume. To the best of our knowledge, this is the first time an anomaly in the LSNO unit cell constants at $T_{\rm CO}$ has been reported.  It correlates well with anomalies in the sound velocity \cite{rami96} and thermal conductivity \cite{hess99}.

We can define an order parameter based on this structural response by quantifying how the $c$/$a$ ratio deviates from a smooth extrapolation of the high-temperature behavior.  We use the $T$-dependence of the unit cell volume for this purpose, which we rescale to lie on top of the $c$/$a$ curve in the high temperature region.   Taking the difference and normalizing the signal to a jump of unity, the resulting
order parameter is shown in the inset of Figure~\ref{fig:LSNO_ca}.  Plotted with it is the normalized single crystal CO peak intensity \cite{du;prl00}, which is proportional to the square of the  conventional order parameter for the charge ordering transition. Both show the same $T$-dependence indicating that the anomaly in $c$/$a$  originates from the development of long-range charge order.

%=====================
%=====================

To understand how charge ordering may couple to the lattice parameters, we consider three extremes: 1) ordered stripes, with a staggered stacking as observed experimentally, and 2) ordered stripes, without any stacking order along the $c$ axis, and 3) uniform charge in layers.  Evaluating the interlayer Coulomb energy for these three configurations, we find that minimization of the Coulomb energy is consistent with an anomalous reduction of $c/a$ ratio at $T_{\rm CO}$.
(The details are presented in the supplementary material.)

\begin{figure}[t!]
\raggedright
\includegraphics[scale=0.77]{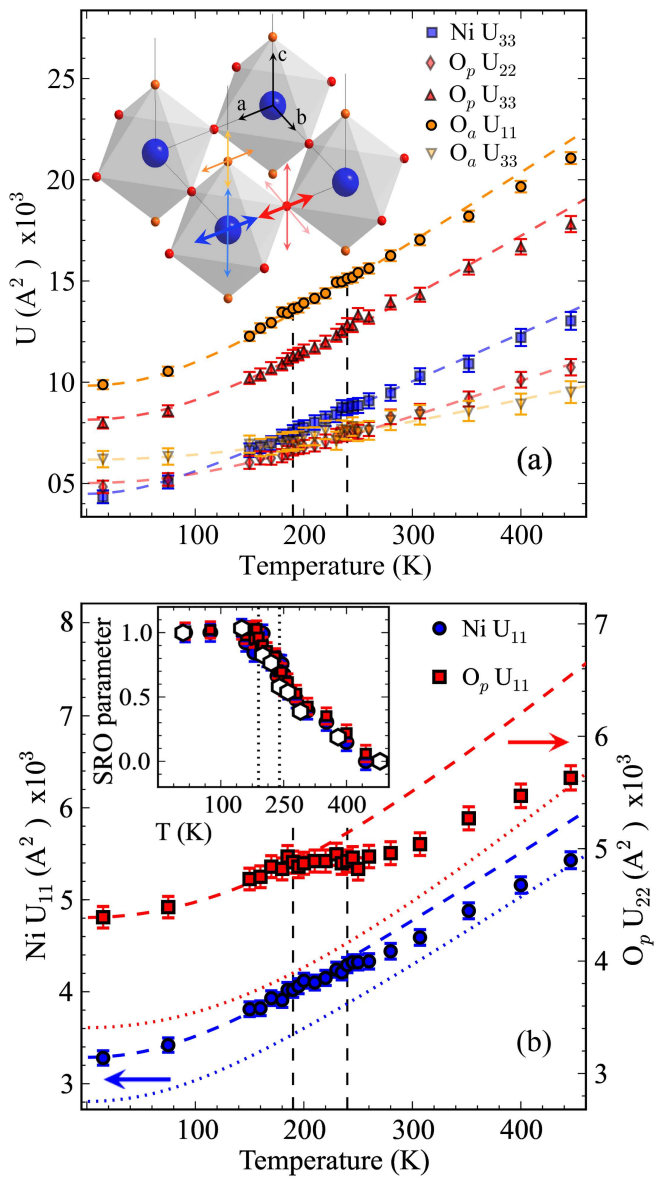}
\caption {(Color online) Anisotropic ADPs with no significant response to CO are shown in (a), while the ADPs showing strong response to CO are shown in (b). Structural motif is shown in the inset of (a), with Ni in blue, apical oxygen (O$_{a}$) in orange and in-plane oxygen (O$_{p}$) in red. Vertical dotted lines denote $T_{\rm SO}$ and $T_{\rm CO}$.  Dashed lines represent calculated Debye-like behavior.  The  dotted lines in (b) represent calculated Debye-like behavior with different offsets from low-T fits. In the inset of (b) order parameters obtained from the ADPs of O$_{p}$~$U_{22}$ and Ni~$U_{11}$ are shown, as well as an order parameter calculated from optical conductivity measurements (open symbols). }
\label{fig:ADP}
\end{figure}

We now turn our attention to the local structure.  Atomic displacements associated with charge stripes, whether static or dynamic, should contribute to measures of disorder relative to the tetragonal model.  From the Rietveld refinement of the diffraction pattern, we obtain atomic displacement parameters (ADPs), where the ADP for a particular atomic site in the unit cell  can be written as $U_{ii} = \langle u_i^2\rangle$, where $u_i$ is the instantaneous displacement of such an atom along direction $i$ and the average is over time and configuration.  In the PDF, disorder will impact the widths of contributions for individual interatomic distributions (although these may overlap and interfere).
For well behaved materials, the temperature dependence of the ADPs is well explained \cite{billi;prb91} by the simple Debye model \cite{debye;adp12}. Deviations from this behavior may indicate the appearance of underlying structural distortions~\cite{billi;prl96,bozin;s10}.  The temperature dependent anisotropic ADPs from various atoms in the LSNO structure obtained from Rietveld refinements are shown in Fig.~\ref{fig:ADP}.
%=========================
%========================
The thermal evolution of various other ADPs are shown in Fig.~\ref{fig:ADP}(a) while the ADPs  lie along in-plane nickel-oxygen bonds are shown in Fig.~\ref{fig:ADP}(b).
In each case the dashed lines (not vertical) represent Debye-model fits to the low-temperature region of the data (with an explanation of the Debye model given in supplementary materials). The directions of the different ADPs are indicated in the inset of \ref{fig:ADP} (a).

For the examples in Fig.~\ref{fig:ADP}(a), the measured ADPs lie on, or close to, the Debye curves at all temperatures, suggesting that the behavior is dominated by conventional disorder due to harmonic motion. The ADPs for oxygen atoms in directions perpendicular to the Ni-O bonds, O$_{a}$$U_{11}$ and O$_{p}$$U_{33}$ are significantly larger in amplitude than the other ADPs in Fig.~\ref{fig:ADP}(a), yielding lower Debye temperatures in the fits, suggesting lower-energy rotational motions of the NiO$_6$ octahedra, as expected in structures based on the perovskite motif, and as noted before.

In contrast, the ADPs for the in-plane O and Ni atoms in directions \emph{parallel} to the in-plane Ni-O bonds, shown in \ref{fig:ADP}(b), exhibit large deviations from conventional behavior.  These ADPs are the ones we anticipate should have substantial contributions associated with stripe order below $T_{\rm CO}$ \cite{tran95b}.  Each ADP should involve a sum of two contributions, one from the stripe-related displacements and another from the usual vibrational motion.  Indeed, we see that the ADPs asymptotically approach a Debye curve at high temperature.  We infer that the excess magnitude of each ADP above the shifted Debye curve (dotted lines) is associated with stripe correlations.  Intriguingly, there is no dramatic change at $T_{\rm CO}$ where static stripe order disappears.  Instead, the low-temperature behavior can be characterized by a distinct Debye curve, with downward deviations beginning near the $T_{\rm SO}$.  The stripe-related disorder above $T_{\rm CO}$ is presumably dynamic, given the effective disappearance of elastic diffuse scattering in single-crystal neutron-scattering experiments.

We can define a measure of local stripe correlations as
%==================================================
\begin{equation}
\label{eq:PDF}
\Delta (T) =\sqrt{U^A_{_{ii}}(T)}  - \sqrt{U_{D_{ii}}^{A}(T)},
\end{equation}
%==================================================
where, $A$ indicates the atomic species and the $U_{D_{ii}}$ represents the Debye approximation to the high temperature behavior.  The ``order'' parameters from in-plane Ni $U_{11}$ and O $U_{22}$, normalized to unity, are shown in the inset of Figure~\ref{fig:ADP}(b).  There is essentially identical temperature dependence for the Ni and O sites.

To make a connection with the electronic properties, one can compare the temperature dependence of the local order parameters with the optical conductivity data from Katsufuji et al.~\cite{katsu;prb96}.
The energy gap for finite conductivity closes at $T_{\rm CO}$, but this provides no measure of the correlations that survive above $T_{\rm CO}$.  To better capture the effective pseudogap, we have evaluated the energy at which the optical conductivity falls to half of its peak value at each temperature.  With normalization at low temperature, this quantity is plotted in the inset of Figure~\ref{fig:ADP}(b) using open symbols. A remarkable resemblance is observed between the local order parameters from the ADPs and the temperature evolution of the energy gap, which directly links the pseudogap behavior in LSNO and the presence of stripe correlations in the local structure.

\begin{figure}%[tb]
\center
\includegraphics[scale=0.74,angle=0]{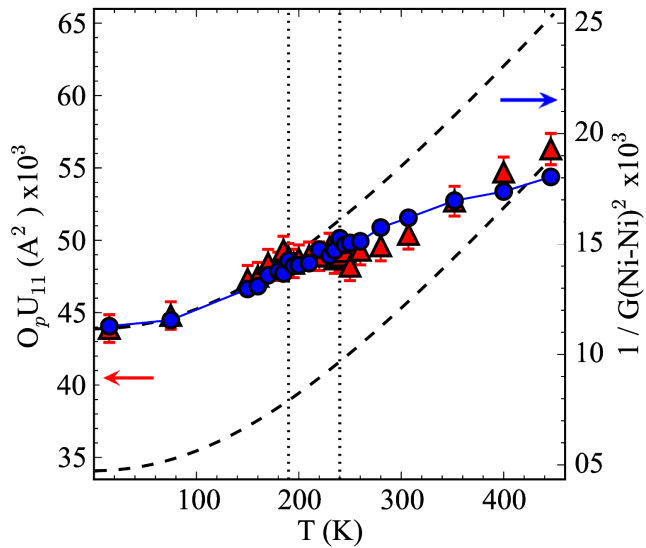}
\caption {(Color online) Comparison of the temperature evolution of the squared inverse height of the nearest neighbor Ni-Ni PDF peak (blue solid circles) and of the in-plane oxygen $U_{22}$. }
\label{fig:PDFhight_ADP}
\end{figure}

The local structural response can also be extracted from PDF analysis.  To keep the analysis simple, we focus on just on the correlations for atoms separated by the $a$ lattice parameter (3.83~\AA), corresponding to the near-neighbor O-O distance that also includes Ni-NI neighbors.  The area of this peak stays essentially constant with temperature, but the width varies.  The inverse square PDF peak height is proportional to the square width, or ADP, of that correlation and so is a sensitive and model independent way of interrogating the data for this parameter.  The temperature evolution of the inverse square PDF peak height is shown in Figure~\ref{fig:PDFhight_ADP} as blue solid circles.
%==========================
%==========================
For comparison, the in-plane oxygen ADP along the Ni-O bond, $U_{22}$, is indicated by red triangles. The inverse square PDF peak height  shows the same $T$ dependence.  A clear break in slope can be observed in the $T$ evolution of the inverse square PDF peak height  at $T_{SO}$ indicating it's sensitivity to stripe order, and the behavior closely follows the Rietveld ADP for the in-plane oxygen along the bond. Thus the PDF which is a model independant complementary measure of the evolution of CO confirms the Rietveld result.

In summary, we have presented a temperature dependent neutron powder diffraction study of LSNO, which is known to show robust charge stripe order below 240~K.  We have shown that the tetragonality shows an anomalous change at $T_{\rm CO}$.  In contrast, a local order parameter determined from mean-square displacements of Ni and in-plane O sites, or equivalently from squared PDF peak widths, shows a temperature dependence that correlates quite well with the pseudogap in the optical conductivity \cite{katsu;prb96}.  From this behavior, we have inferred that charge-stripe correlations survive to temperatures far above $T_{\rm CO}$, and only gradually become attenuated.  These results provide support for the idea that fluctuating stripes could underlie electronic nematic behavior in cuprates \cite{kive98,lawl10}.

This work was supported by the Office of Basic Energy Sciences, Division of Materials Sciences and Engineering, U.S. Department of Energy through account DE-AC02-98CH10886. This work has benefited from the use of NPDF at LANSCE, funded by DOE Office of Basic Energy
Sciences. LANL is operated by Los Alamos National Security LLC under DOE Contract No. DE-AC52-06NA25396.

%\bibliography{billinge-group,abb-billinge-group,everyone,LNO,theory}

\begin{thebibliography}{40}%
\makeatletter
\providecommand \@ifxundefined [1]{%
 \@ifx{#1\undefined}
}%
\providecommand \@ifnum [1]{%
 \ifnum #1\expandafter \@firstoftwo
 \else \expandafter \@secondoftwo
 \fi
}%
\providecommand \@ifx [1]{%
 \ifx #1\expandafter \@firstoftwo
 \else \expandafter \@secondoftwo
 \fi
}%
\providecommand \natexlab [1]{#1}%
\providecommand \enquote  [1]{``#1''}%
\providecommand \bibnamefont  [1]{#1}%
\providecommand \bibfnamefont [1]{#1}%
\providecommand \citenamefont [1]{#1}%
\providecommand \href@noop [0]{\@secondoftwo}%
\providecommand \href [0]{\begingroup \@sanitize@url \@href}%
\providecommand \@href[1]{\@@startlink{#1}\@@href}%
\providecommand \@@href[1]{\endgroup#1\@@endlink}%
\providecommand \@sanitize@url [0]{\catcode `\\12\catcode `\$12\catcode
  `\&12\catcode `\#12\catcode `\^12\catcode `\_12\catcode `\%12\relax}%
\providecommand \@@startlink[1]{}%
\providecommand \@@endlink[0]{}%
\providecommand \url  [0]{\begingroup\@sanitize@url \@url }%
\providecommand \@url [1]{\endgroup\@href {#1}{\urlprefix }}%
\providecommand \urlprefix  [0]{URL }%
\providecommand \Eprint [0]{\href }%
\@ifxundefined \urlstyle {%
  \providecommand \doi  [0]{\begingroup \@sanitize@url \@doi}%
  \providecommand \@doi [1]{\endgroup \@@startlink {\doibase
  #1}doi:\discretionary {}{}{}#1\@@endlink }%
}{%
  \providecommand \doi  [0]{doi:\discretionary{}{}{}\begingroup
  \urlstyle{rm}\Url }%
}%
\providecommand \doibase [0]{http://dx.doi.org/}%
\providecommand \Doi [0]{\begingroup \@sanitize@url \@Doi }%
\providecommand \@Doi  [1]{\endgroup\@@startlink{\doibase#1}\@@Doi}%
\providecommand \@@Doi [1]{#1\@@endlink}%
\providecommand \selectlanguage [0]{\@gobble}%
\providecommand \bibinfo  [0]{\@secondoftwo}%
\providecommand \bibfield  [0]{\@secondoftwo}%
\providecommand \translation [1]{[#1]}%
\providecommand \BibitemOpen [0]{}%
\providecommand \bibitemStop [0]{}%
\providecommand \bibitemNoStop [0]{.\EOS\space}%
\providecommand \EOS [0]{\spacefactor3000\relax}%
\providecommand \BibitemShut  [1]{\csname bibitem#1\endcsname}%
%</preamble>
\bibitem [{\citenamefont {Kivelson}\ \emph {et~al.}(2003)\citenamefont
  {Kivelson}, \citenamefont {Bindloss}, \citenamefont {Fradkin}, \citenamefont
  {Oganesyan}, \citenamefont {Tranquada}, \citenamefont {Kapitulnik},\ and\
  \citenamefont {Howald}}]{kive03}%
  \BibitemOpen
  \bibfield  {author} {\bibinfo {author} {\bibfnamefont {S.~A.}\ \bibnamefont
  {Kivelson}}, \bibinfo {author} {\bibfnamefont {I.~P.}\ \bibnamefont
  {Bindloss}}, \bibinfo {author} {\bibfnamefont {E.}~\bibnamefont {Fradkin}},
  \bibinfo {author} {\bibfnamefont {V.}~\bibnamefont {Oganesyan}}, \bibinfo
  {author} {\bibfnamefont {J.~M.}\ \bibnamefont {Tranquada}}, \bibinfo {author}
  {\bibfnamefont {A.}~\bibnamefont {Kapitulnik}}, \ and\ \bibinfo {author}
  {\bibfnamefont {C.}~\bibnamefont {Howald}},\ }\href@noop {} {\bibfield
  {journal} {\bibinfo  {journal} {Rev. Mod. Phys.},\ }\textbf {\bibinfo
  {volume} {75}},\ \bibinfo {pages} {1201} (\bibinfo {year}
  {2003})}\BibitemShut {NoStop}%
\bibitem [{\citenamefont {Vojta}(2009)}]{vojt09}%
  \BibitemOpen
  \bibfield  {author} {\bibinfo {author} {\bibfnamefont {M.}~\bibnamefont
  {Vojta}},\ }\href@noop {} {\bibfield  {journal} {\bibinfo  {journal} {Adv.
  Phys.},\ }\textbf {\bibinfo {volume} {58}},\ \bibinfo {pages} {699} (\bibinfo
  {year} {2009})}\BibitemShut {NoStop}%
\bibitem [{\citenamefont {Zaanen}\ \emph {et~al.}(2001)\citenamefont {Zaanen},
  \citenamefont {Osman}, \citenamefont {Kruis}, \citenamefont {Nussinov},\ and\
  \citenamefont {Tworzyd{\l}o}}]{zaan01}%
  \BibitemOpen
  \bibfield  {author} {\bibinfo {author} {\bibfnamefont {J.}~\bibnamefont
  {Zaanen}}, \bibinfo {author} {\bibfnamefont {O.~Y.}\ \bibnamefont {Osman}},
  \bibinfo {author} {\bibfnamefont {H.~V.}\ \bibnamefont {Kruis}}, \bibinfo
  {author} {\bibfnamefont {Z.}~\bibnamefont {Nussinov}}, \ and\ \bibinfo
  {author} {\bibfnamefont {J.}~\bibnamefont {Tworzyd{\l}o}},\ }\href@noop {}
  {\bibfield  {journal} {\bibinfo  {journal} {Phil. Mag. B},\ }\textbf
  {\bibinfo {volume} {81}},\ \bibinfo {pages} {1485} (\bibinfo {year}
  {2001})}\BibitemShut {NoStop}%
\bibitem [{\citenamefont {H\"ucker}(2012)}]{huck12}%
  \BibitemOpen
  \bibfield  {author} {\bibinfo {author} {\bibfnamefont {M.}~\bibnamefont
  {H\"ucker}},\ }\href@noop {} {\bibfield  {journal} {\bibinfo  {journal}
  {Physica C},\ }\textbf {\bibinfo {volume} {481}},\ \bibinfo {pages} {3}
  (\bibinfo {year} {2012})}\BibitemShut {NoStop}%
\bibitem [{\citenamefont {Abbamonte}\ \emph {et~al.}(2012)\citenamefont
  {Abbamonte}, \citenamefont {Demler}, \citenamefont {Davis},\ and\
  \citenamefont {Campuzano}}]{abba12}%
  \BibitemOpen
  \bibfield  {author} {\bibinfo {author} {\bibfnamefont {P.}~\bibnamefont
  {Abbamonte}}, \bibinfo {author} {\bibfnamefont {E.}~\bibnamefont {Demler}},
  \bibinfo {author} {\bibfnamefont {J.~S.}\ \bibnamefont {Davis}}, \ and\
  \bibinfo {author} {\bibfnamefont {J.-C.}\ \bibnamefont {Campuzano}},\
  }\href@noop {} {\bibfield  {journal} {\bibinfo  {journal} {Physica C},\
  }\textbf {\bibinfo {volume} {481}},\ \bibinfo {pages} {15} (\bibinfo {year}
  {2012})}\BibitemShut {NoStop}%
\bibitem [{\citenamefont {Fujita}(2012)}]{fuji12b}%
  \BibitemOpen
  \bibfield  {author} {\bibinfo {author} {\bibfnamefont {M.}~\bibnamefont
  {Fujita}},\ }\href@noop {} {\bibfield  {journal} {\bibinfo  {journal}
  {Physica C},\ }\textbf {\bibinfo {volume} {481}},\ \bibinfo {pages} {23}
  (\bibinfo {year} {2012})}\BibitemShut {NoStop}%
\bibitem [{\citenamefont {Robertson}\ \emph {et~al.}(2006)\citenamefont
  {Robertson}, \citenamefont {Kivelson}, \citenamefont {Fradkin}, \citenamefont
  {Fang},\ and\ \citenamefont {Kapitulnik}}]{robe06}%
  \BibitemOpen
  \bibfield  {author} {\bibinfo {author} {\bibfnamefont {J.~A.}\ \bibnamefont
  {Robertson}}, \bibinfo {author} {\bibfnamefont {S.~A.}\ \bibnamefont
  {Kivelson}}, \bibinfo {author} {\bibfnamefont {E.}~\bibnamefont {Fradkin}},
  \bibinfo {author} {\bibfnamefont {A.~C.}\ \bibnamefont {Fang}}, \ and\
  \bibinfo {author} {\bibfnamefont {A.}~\bibnamefont {Kapitulnik}},\ }\Doi
  {10.1103/PhysRevB.74.134507} {\bibfield  {journal} {\bibinfo  {journal}
  {Phys. Rev. B},\ }\textbf {\bibinfo {volume} {74}},\ \bibinfo {pages}
  {134507} (\bibinfo {year} {2006})}\BibitemShut {NoStop}%
\bibitem [{\citenamefont {Kohsaka}\ \emph {et~al.}(2007)\citenamefont
  {Kohsaka}, \citenamefont {Taylor}, \citenamefont {Fujita}, \citenamefont
  {Schmidt}, \citenamefont {Lupien}, \citenamefont {Hanaguri}, \citenamefont
  {Azuma}, \citenamefont {Takano}, \citenamefont {Eisaki}, \citenamefont
  {Takagi}, \citenamefont {Uchida},\ and\ \citenamefont {Davis}}]{kohs07}%
  \BibitemOpen
  \bibfield  {author} {\bibinfo {author} {\bibfnamefont {Y.}~\bibnamefont
  {Kohsaka}}, \bibinfo {author} {\bibfnamefont {C.}~\bibnamefont {Taylor}},
  \bibinfo {author} {\bibfnamefont {K.}~\bibnamefont {Fujita}}, \bibinfo
  {author} {\bibfnamefont {A.}~\bibnamefont {Schmidt}}, \bibinfo {author}
  {\bibfnamefont {C.}~\bibnamefont {Lupien}}, \bibinfo {author} {\bibfnamefont
  {T.}~\bibnamefont {Hanaguri}}, \bibinfo {author} {\bibfnamefont
  {M.}~\bibnamefont {Azuma}}, \bibinfo {author} {\bibfnamefont
  {M.}~\bibnamefont {Takano}}, \bibinfo {author} {\bibfnamefont
  {H.}~\bibnamefont {Eisaki}}, \bibinfo {author} {\bibfnamefont
  {H.}~\bibnamefont {Takagi}}, \bibinfo {author} {\bibfnamefont
  {S.}~\bibnamefont {Uchida}}, \ and\ \bibinfo {author} {\bibfnamefont {J.~C.}\
  \bibnamefont {Davis}},\ }\href@noop {} {\bibfield  {journal} {\bibinfo
  {journal} {Science},\ }\textbf {\bibinfo {volume} {315}},\ \bibinfo {pages}
  {1380} (\bibinfo {year} {2007})}\BibitemShut {NoStop}%
\bibitem [{\citenamefont {Parker}\ \emph {et~al.}(2010)\citenamefont {Parker},
  \citenamefont {Aynajian}, \citenamefont {da~Silva~Neto}, \citenamefont
  {Pushp}, \citenamefont {Ono}, \citenamefont {Wen}, \citenamefont {Xu},
  \citenamefont {Gu},\ and\ \citenamefont {Yazdani}}]{park10}%
  \BibitemOpen
  \bibfield  {author} {\bibinfo {author} {\bibfnamefont {C.~V.}\ \bibnamefont
  {Parker}}, \bibinfo {author} {\bibfnamefont {P.}~\bibnamefont {Aynajian}},
  \bibinfo {author} {\bibfnamefont {E.~H.}\ \bibnamefont {da~Silva~Neto}},
  \bibinfo {author} {\bibfnamefont {A.}~\bibnamefont {Pushp}}, \bibinfo
  {author} {\bibfnamefont {S.}~\bibnamefont {Ono}}, \bibinfo {author}
  {\bibfnamefont {J.}~\bibnamefont {Wen}}, \bibinfo {author} {\bibfnamefont
  {Z.}~\bibnamefont {Xu}}, \bibinfo {author} {\bibfnamefont {G.}~\bibnamefont
  {Gu}}, \ and\ \bibinfo {author} {\bibfnamefont {A.}~\bibnamefont {Yazdani}},\
  }\href@noop {} {\bibfield  {journal} {\bibinfo  {journal} {Nature},\ }\textbf
  {\bibinfo {volume} {468}},\ \bibinfo {pages} {677} (\bibinfo {year}
  {2010})}\BibitemShut {NoStop}%
\bibitem [{\citenamefont {Ghiringhelli}\ \emph {et~al.}(2012)\citenamefont
  {Ghiringhelli}, \citenamefont {Le~Tacon}, \citenamefont {Minola},
  \citenamefont {Blanco-Canosa}, \citenamefont {Mazzoli}, \citenamefont
  {Brookes}, \citenamefont {De~Luca}, \citenamefont {Frano}, \citenamefont
  {Hawthorn}, \citenamefont {He}, \citenamefont {Loew}, \citenamefont {Sala},
  \citenamefont {Peets}, \citenamefont {Salluzzo}, \citenamefont {Schierle},
  \citenamefont {Sutarto}, \citenamefont {Sawatzky}, \citenamefont {Weschke},
  \citenamefont {Keimer},\ and\ \citenamefont {Braicovich}}]{ghir12}%
  \BibitemOpen
  \bibfield  {author} {\bibinfo {author} {\bibfnamefont {G.}~\bibnamefont
  {Ghiringhelli}}, \bibinfo {author} {\bibfnamefont {M.}~\bibnamefont
  {Le~Tacon}}, \bibinfo {author} {\bibfnamefont {M.}~\bibnamefont {Minola}},
  \bibinfo {author} {\bibfnamefont {S.}~\bibnamefont {Blanco-Canosa}}, \bibinfo
  {author} {\bibfnamefont {C.}~\bibnamefont {Mazzoli}}, \bibinfo {author}
  {\bibfnamefont {N.~B.}\ \bibnamefont {Brookes}}, \bibinfo {author}
  {\bibfnamefont {G.~M.}\ \bibnamefont {De~Luca}}, \bibinfo {author}
  {\bibfnamefont {A.}~\bibnamefont {Frano}}, \bibinfo {author} {\bibfnamefont
  {D.~G.}\ \bibnamefont {Hawthorn}}, \bibinfo {author} {\bibfnamefont
  {F.}~\bibnamefont {He}}, \bibinfo {author} {\bibfnamefont {T.}~\bibnamefont
  {Loew}}, \bibinfo {author} {\bibfnamefont {M.~M.}\ \bibnamefont {Sala}},
  \bibinfo {author} {\bibfnamefont {D.~C.}\ \bibnamefont {Peets}}, \bibinfo
  {author} {\bibfnamefont {M.}~\bibnamefont {Salluzzo}}, \bibinfo {author}
  {\bibfnamefont {E.}~\bibnamefont {Schierle}}, \bibinfo {author}
  {\bibfnamefont {R.}~\bibnamefont {Sutarto}}, \bibinfo {author} {\bibfnamefont
  {G.~A.}\ \bibnamefont {Sawatzky}}, \bibinfo {author} {\bibfnamefont
  {E.}~\bibnamefont {Weschke}}, \bibinfo {author} {\bibfnamefont
  {B.}~\bibnamefont {Keimer}}, \ and\ \bibinfo {author} {\bibfnamefont
  {L.}~\bibnamefont {Braicovich}},\ }\href@noop {} {\bibfield  {journal}
  {\bibinfo  {journal} {Science},\ }\textbf {\bibinfo {volume} {337}},\
  \bibinfo {pages} {821} (\bibinfo {year} {2012})}\BibitemShut {NoStop}%
\bibitem [{\citenamefont {Chang}\ \emph {et~al.}(2012)\citenamefont {Chang},
  \citenamefont {Blackburn}, \citenamefont {Holmes}, \citenamefont
  {Christensen}, \citenamefont {Larsen}, \citenamefont {Mesot}, \citenamefont
  {Liang}, \citenamefont {Bonn}, \citenamefont {Hardy}, \citenamefont
  {Watenphul}, \citenamefont {Zimmermann}, \citenamefont {Forgan},\ and\
  \citenamefont {Hayden}}]{chan12a}%
  \BibitemOpen
  \bibfield  {author} {\bibinfo {author} {\bibfnamefont {J.}~\bibnamefont
  {Chang}}, \bibinfo {author} {\bibfnamefont {E.}~\bibnamefont {Blackburn}},
  \bibinfo {author} {\bibfnamefont {A.~T.}\ \bibnamefont {Holmes}}, \bibinfo
  {author} {\bibfnamefont {N.~B.}\ \bibnamefont {Christensen}}, \bibinfo
  {author} {\bibfnamefont {J.}~\bibnamefont {Larsen}}, \bibinfo {author}
  {\bibfnamefont {J.}~\bibnamefont {Mesot}}, \bibinfo {author} {\bibfnamefont
  {R.}~\bibnamefont {Liang}}, \bibinfo {author} {\bibfnamefont {D.~A.}\
  \bibnamefont {Bonn}}, \bibinfo {author} {\bibfnamefont {W.~N.}\ \bibnamefont
  {Hardy}}, \bibinfo {author} {\bibfnamefont {A.}~\bibnamefont {Watenphul}},
  \bibinfo {author} {\bibfnamefont {M.~v.}\ \bibnamefont {Zimmermann}},
  \bibinfo {author} {\bibfnamefont {E.~M.}\ \bibnamefont {Forgan}}, \ and\
  \bibinfo {author} {\bibfnamefont {S.~M.}\ \bibnamefont {Hayden}},\
  }\href@noop {} {\bibfield  {journal} {\bibinfo  {journal} {Nat. Phys.},\
  }\textbf {\bibinfo {volume} {8}},\ \bibinfo {pages} {871} (\bibinfo {year}
  {2012})}\BibitemShut {NoStop}%
\bibitem [{\citenamefont {Blackburn}\ \emph {et~al.}(2013)\citenamefont
  {Blackburn}, \citenamefont {Chang}, \citenamefont {H\"ucker}, \citenamefont
  {Holmes}, \citenamefont {Christensen}, \citenamefont {Liang}, \citenamefont
  {Bonn}, \citenamefont {Hardy}, \citenamefont {R\"utt}, \citenamefont
  {Gutowski}, \citenamefont {Zimmermann}, \citenamefont {Forgan},\ and\
  \citenamefont {Hayden}}]{blac13}%
  \BibitemOpen
  \bibfield  {author} {\bibinfo {author} {\bibfnamefont {E.}~\bibnamefont
  {Blackburn}}, \bibinfo {author} {\bibfnamefont {J.}~\bibnamefont {Chang}},
  \bibinfo {author} {\bibfnamefont {M.}~\bibnamefont {H\"ucker}}, \bibinfo
  {author} {\bibfnamefont {A.~T.}\ \bibnamefont {Holmes}}, \bibinfo {author}
  {\bibfnamefont {N.~B.}\ \bibnamefont {Christensen}}, \bibinfo {author}
  {\bibfnamefont {R.}~\bibnamefont {Liang}}, \bibinfo {author} {\bibfnamefont
  {D.~A.}\ \bibnamefont {Bonn}}, \bibinfo {author} {\bibfnamefont {W.~N.}\
  \bibnamefont {Hardy}}, \bibinfo {author} {\bibfnamefont {U.}~\bibnamefont
  {R\"utt}}, \bibinfo {author} {\bibfnamefont {O.}~\bibnamefont {Gutowski}},
  \bibinfo {author} {\bibfnamefont {M.~v.}\ \bibnamefont {Zimmermann}},
  \bibinfo {author} {\bibfnamefont {E.~M.}\ \bibnamefont {Forgan}}, \ and\
  \bibinfo {author} {\bibfnamefont {S.~M.}\ \bibnamefont {Hayden}},\ }\Doi
  {10.1103/PhysRevLett.110.137004} {\bibfield  {journal} {\bibinfo  {journal}
  {Phys. Rev. Lett.},\ }\textbf {\bibinfo {volume} {110}},\ \bibinfo {pages}
  {137004} (\bibinfo {year} {2013})}\BibitemShut {NoStop}%
\bibitem [{\citenamefont {Blanco-Canosa}\ \emph {et~al.}(2013)\citenamefont
  {Blanco-Canosa}, \citenamefont {Frano}, \citenamefont {Loew}, \citenamefont
  {Lu}, \citenamefont {Porras}, \citenamefont {Ghiringhelli}, \citenamefont
  {Minola}, \citenamefont {Mazzoli}, \citenamefont {Braicovich}, \citenamefont
  {Schierle}, \citenamefont {Weschke}, \citenamefont {Le~Tacon},\ and\
  \citenamefont {Keimer}}]{blan13}%
  \BibitemOpen
  \bibfield  {author} {\bibinfo {author} {\bibfnamefont {S.}~\bibnamefont
  {Blanco-Canosa}}, \bibinfo {author} {\bibfnamefont {A.}~\bibnamefont
  {Frano}}, \bibinfo {author} {\bibfnamefont {T.}~\bibnamefont {Loew}},
  \bibinfo {author} {\bibfnamefont {Y.}~\bibnamefont {Lu}}, \bibinfo {author}
  {\bibfnamefont {J.}~\bibnamefont {Porras}}, \bibinfo {author} {\bibfnamefont
  {G.}~\bibnamefont {Ghiringhelli}}, \bibinfo {author} {\bibfnamefont
  {M.}~\bibnamefont {Minola}}, \bibinfo {author} {\bibfnamefont
  {C.}~\bibnamefont {Mazzoli}}, \bibinfo {author} {\bibfnamefont
  {L.}~\bibnamefont {Braicovich}}, \bibinfo {author} {\bibfnamefont
  {E.}~\bibnamefont {Schierle}}, \bibinfo {author} {\bibfnamefont
  {E.}~\bibnamefont {Weschke}}, \bibinfo {author} {\bibfnamefont
  {M.}~\bibnamefont {Le~Tacon}}, \ and\ \bibinfo {author} {\bibfnamefont
  {B.}~\bibnamefont {Keimer}},\ }\Doi {10.1103/PhysRevLett.110.187001}
  {\bibfield  {journal} {\bibinfo  {journal} {Phys. Rev. Lett.},\ }\textbf
  {\bibinfo {volume} {110}},\ \bibinfo {pages} {187001} (\bibinfo {year}
  {2013})}\BibitemShut {NoStop}%
\bibitem [{\citenamefont {{Thampy}}\ \emph {et~al.}(2013)\citenamefont
  {{Thampy}}, \citenamefont {{Blanco-Canosa}}, \citenamefont
  {{Garc{\'{\i}}a-Fern{\'a}ndez}}, \citenamefont {{Dean}}, \citenamefont
  {{Gu}}, \citenamefont {{F{\"o}erst}}, \citenamefont {{Keimer}}, \citenamefont
  {{Le Tacon}}, \citenamefont {{Wilkins}},\ and\ \citenamefont
  {{Hill}}}]{2013arXiv1305.5515T}%
  \BibitemOpen
  \bibfield  {author} {\bibinfo {author} {\bibfnamefont {V.}~\bibnamefont
  {{Thampy}}}, \bibinfo {author} {\bibfnamefont {S.}~\bibnamefont
  {{Blanco-Canosa}}}, \bibinfo {author} {\bibfnamefont {M.}~\bibnamefont
  {{Garc{\'{\i}}a-Fern{\'a}ndez}}}, \bibinfo {author} {\bibfnamefont
  {M.~P.~M.}\ \bibnamefont {{Dean}}}, \bibinfo {author} {\bibfnamefont {G.~D.}\
  \bibnamefont {{Gu}}}, \bibinfo {author} {\bibfnamefont {M.}~\bibnamefont
  {{F{\"o}erst}}}, \bibinfo {author} {\bibfnamefont {B.}~\bibnamefont
  {{Keimer}}}, \bibinfo {author} {\bibfnamefont {M.}~\bibnamefont {{Le
  Tacon}}}, \bibinfo {author} {\bibfnamefont {S.~B.}\ \bibnamefont
  {{Wilkins}}}, \ and\ \bibinfo {author} {\bibfnamefont {J.~P.}\ \bibnamefont
  {{Hill}}},\ }\href@noop {} {\bibfield  {journal} {\bibinfo  {journal} {ArXiv
  e-prints}} (\bibinfo {year} {2013})},\ \Eprint
  {http://arxiv.org/abs/1305.5515} {arXiv:1305.5515 [cond-mat.supr-con]}
  \BibitemShut {NoStop}%
\bibitem [{\citenamefont {H\"ucker}\ \emph {et~al.}(2011)\citenamefont
  {H\"ucker}, \citenamefont {v.~Zimmermann}, \citenamefont {Gu}, \citenamefont
  {Xu}, \citenamefont {Wen}, \citenamefont {Xu}, \citenamefont {Kang},
  \citenamefont {Zheludev},\ and\ \citenamefont {Tranquada}}]{huck11}%
  \BibitemOpen
  \bibfield  {author} {\bibinfo {author} {\bibfnamefont {M.}~\bibnamefont
  {H\"ucker}}, \bibinfo {author} {\bibfnamefont {M.}~\bibnamefont
  {v.~Zimmermann}}, \bibinfo {author} {\bibfnamefont {G.~D.}\ \bibnamefont
  {Gu}}, \bibinfo {author} {\bibfnamefont {Z.~J.}\ \bibnamefont {Xu}}, \bibinfo
  {author} {\bibfnamefont {J.~S.}\ \bibnamefont {Wen}}, \bibinfo {author}
  {\bibfnamefont {G.}~\bibnamefont {Xu}}, \bibinfo {author} {\bibfnamefont
  {H.~J.}\ \bibnamefont {Kang}}, \bibinfo {author} {\bibfnamefont
  {A.}~\bibnamefont {Zheludev}}, \ and\ \bibinfo {author} {\bibfnamefont
  {J.~M.}\ \bibnamefont {Tranquada}},\ }\href@noop {} {\bibfield  {journal}
  {\bibinfo  {journal} {Phys. Rev. B},\ }\textbf {\bibinfo {volume} {83}},\
  \bibinfo {pages} {104506} (\bibinfo {year} {2011})}\BibitemShut {NoStop}%
\bibitem [{\citenamefont {Tajima}\ \emph {et~al.}(2001)\citenamefont {Tajima},
  \citenamefont {Noda}, \citenamefont {Eisaki},\ and\ \citenamefont
  {Uchida}}]{taji01}%
  \BibitemOpen
  \bibfield  {author} {\bibinfo {author} {\bibfnamefont {S.}~\bibnamefont
  {Tajima}}, \bibinfo {author} {\bibfnamefont {T.}~\bibnamefont {Noda}},
  \bibinfo {author} {\bibfnamefont {H.}~\bibnamefont {Eisaki}}, \ and\ \bibinfo
  {author} {\bibfnamefont {S.}~\bibnamefont {Uchida}},\ }\href@noop {}
  {\bibfield  {journal} {\bibinfo  {journal} {Phys. Rev. Lett.},\ }\textbf
  {\bibinfo {volume} {86}},\ \bibinfo {pages} {500} (\bibinfo {year}
  {2001})}\BibitemShut {NoStop}%
\bibitem [{\citenamefont {Li}\ \emph {et~al.}(2007)\citenamefont {Li},
  \citenamefont {{H\"ucker}}, \citenamefont {Gu}, \citenamefont {Tsvelik},\
  and\ \citenamefont {Tranquada}}]{li07}%
  \BibitemOpen
  \bibfield  {author} {\bibinfo {author} {\bibfnamefont {Q.}~\bibnamefont
  {Li}}, \bibinfo {author} {\bibfnamefont {M.}~\bibnamefont {{H\"ucker}}},
  \bibinfo {author} {\bibfnamefont {G.~D.}\ \bibnamefont {Gu}}, \bibinfo
  {author} {\bibfnamefont {A.~M.}\ \bibnamefont {Tsvelik}}, \ and\ \bibinfo
  {author} {\bibfnamefont {J.~M.}\ \bibnamefont {Tranquada}},\ }\href@noop {}
  {\bibfield  {journal} {\bibinfo  {journal} {Phys. Rev. Lett.},\ }\textbf
  {\bibinfo {volume} {99}},\ \bibinfo {eid} {067001} (\bibinfo {year}
  {2007})}\BibitemShut {NoStop}%
\bibitem [{\citenamefont {Berg}\ \emph {et~al.}(2009)\citenamefont {Berg},
  \citenamefont {Fradkin}, \citenamefont {Kivelson},\ and\ \citenamefont
  {Tranquada}}]{berg09b}%
  \BibitemOpen
  \bibfield  {author} {\bibinfo {author} {\bibfnamefont {E.}~\bibnamefont
  {Berg}}, \bibinfo {author} {\bibfnamefont {E.}~\bibnamefont {Fradkin}},
  \bibinfo {author} {\bibfnamefont {S.~A.}\ \bibnamefont {Kivelson}}, \ and\
  \bibinfo {author} {\bibfnamefont {J.~M.}\ \bibnamefont {Tranquada}},\
  }\href@noop {} {\bibfield  {journal} {\bibinfo  {journal} {New J. Phys.},\
  }\textbf {\bibinfo {volume} {11}},\ \bibinfo {pages} {115004} (\bibinfo
  {year} {2009})}\BibitemShut {NoStop}%
\bibitem [{\citenamefont {Homes}\ \emph {et~al.}(2012)\citenamefont {Homes},
  \citenamefont {H\"ucker}, \citenamefont {Li}, \citenamefont {Xu},
  \citenamefont {Wen}, \citenamefont {Gu},\ and\ \citenamefont
  {Tranquada}}]{home12}%
  \BibitemOpen
  \bibfield  {author} {\bibinfo {author} {\bibfnamefont {C.~C.}\ \bibnamefont
  {Homes}}, \bibinfo {author} {\bibfnamefont {M.}~\bibnamefont {H\"ucker}},
  \bibinfo {author} {\bibfnamefont {Q.}~\bibnamefont {Li}}, \bibinfo {author}
  {\bibfnamefont {Z.~J.}\ \bibnamefont {Xu}}, \bibinfo {author} {\bibfnamefont
  {J.~S.}\ \bibnamefont {Wen}}, \bibinfo {author} {\bibfnamefont {G.~D.}\
  \bibnamefont {Gu}}, \ and\ \bibinfo {author} {\bibfnamefont {J.~M.}\
  \bibnamefont {Tranquada}},\ }\href@noop {} {\bibfield  {journal} {\bibinfo
  {journal} {Phys. Rev. B},\ }\textbf {\bibinfo {volume} {85}},\ \bibinfo
  {pages} {134510} (\bibinfo {year} {2012})}\BibitemShut {NoStop}%
\bibitem [{\citenamefont {Kivelson}\ \emph {et~al.}(1998)\citenamefont
  {Kivelson}, \citenamefont {Fradkin},\ and\ \citenamefont {Emery}}]{kive98}%
  \BibitemOpen
  \bibfield  {author} {\bibinfo {author} {\bibfnamefont {S.~A.}\ \bibnamefont
  {Kivelson}}, \bibinfo {author} {\bibfnamefont {E.}~\bibnamefont {Fradkin}}, \
  and\ \bibinfo {author} {\bibfnamefont {V.~J.}\ \bibnamefont {Emery}},\
  }\href@noop {} {\bibfield  {journal} {\bibinfo  {journal} {Nature},\ }\textbf
  {\bibinfo {volume} {393}},\ \bibinfo {pages} {550} (\bibinfo {year}
  {1998})}\BibitemShut {NoStop}%
\bibitem [{\citenamefont {Ulbrich}\ and\ \citenamefont
  {Braden}(2012)}]{ulbr12b}%
  \BibitemOpen
  \bibfield  {author} {\bibinfo {author} {\bibfnamefont {H.}~\bibnamefont
  {Ulbrich}}\ and\ \bibinfo {author} {\bibfnamefont {M.}~\bibnamefont
  {Braden}},\ }\Doi {10.1016/j.physc.2012.04.039} {\bibfield  {journal}
  {\bibinfo  {journal} {Physica C},\ }\textbf {\bibinfo {volume} {481}},\
  \bibinfo {pages} {31} (\bibinfo {year} {2012})}\BibitemShut {NoStop}%
\bibitem [{\citenamefont {Ramirez}\ \emph {et~al.}(1996)\citenamefont
  {Ramirez}, \citenamefont {Gammel}, \citenamefont {Cheong}, \citenamefont
  {Bishop},\ and\ \citenamefont {Chandra}}]{rami96}%
  \BibitemOpen
  \bibfield  {author} {\bibinfo {author} {\bibfnamefont {A.~P.}\ \bibnamefont
  {Ramirez}}, \bibinfo {author} {\bibfnamefont {P.~L.}\ \bibnamefont {Gammel}},
  \bibinfo {author} {\bibfnamefont {S.-W.}\ \bibnamefont {Cheong}}, \bibinfo
  {author} {\bibfnamefont {D.~J.}\ \bibnamefont {Bishop}}, \ and\ \bibinfo
  {author} {\bibfnamefont {P.}~\bibnamefont {Chandra}},\ }\href@noop {}
  {\bibfield  {journal} {\bibinfo  {journal} {Phys. Rev. Lett.},\ }\textbf
  {\bibinfo {volume} {76}},\ \bibinfo {pages} {447} (\bibinfo {year}
  {1996})}\BibitemShut {NoStop}%
\bibitem [{\citenamefont {Lee}\ and\ \citenamefont {Cheong}(1997)}]{lee97}%
  \BibitemOpen
  \bibfield  {author} {\bibinfo {author} {\bibfnamefont {S.-H.}\ \bibnamefont
  {Lee}}\ and\ \bibinfo {author} {\bibfnamefont {S.-W.}\ \bibnamefont
  {Cheong}},\ }\href@noop {} {\bibfield  {journal} {\bibinfo  {journal} {Phys.
  Rev. Lett.},\ }\textbf {\bibinfo {volume} {79}},\ \bibinfo {pages} {2514}
  (\bibinfo {year} {1997})}\BibitemShut {NoStop}%
\bibitem [{\citenamefont {Lee}\ \emph {et~al.}(2002)\citenamefont {Lee},
  \citenamefont {Tranquada}, \citenamefont {Yamada}, \citenamefont {Buttrey},
  \citenamefont {Li},\ and\ \citenamefont {Cheong}}]{lee02}%
  \BibitemOpen
  \bibfield  {author} {\bibinfo {author} {\bibfnamefont {S.-H.}\ \bibnamefont
  {Lee}}, \bibinfo {author} {\bibfnamefont {J.~M.}\ \bibnamefont {Tranquada}},
  \bibinfo {author} {\bibfnamefont {K.}~\bibnamefont {Yamada}}, \bibinfo
  {author} {\bibfnamefont {D.~J.}\ \bibnamefont {Buttrey}}, \bibinfo {author}
  {\bibfnamefont {Q.}~\bibnamefont {Li}}, \ and\ \bibinfo {author}
  {\bibfnamefont {S.-W.}\ \bibnamefont {Cheong}},\ }\href@noop {} {\bibfield
  {journal} {\bibinfo  {journal} {Phys. Rev. Lett.},\ }\textbf {\bibinfo
  {volume} {88}},\ \bibinfo {pages} {126401} (\bibinfo {year}
  {2002})}\BibitemShut {NoStop}%
\bibitem [{\citenamefont {Du}\ \emph {et~al.}(2000)\citenamefont {Du},
  \citenamefont {Ghazi}, \citenamefont {Su}, \citenamefont {Pape},\ and\
  \citenamefont {Hatton}}]{du;prl00}%
  \BibitemOpen
  \bibfield  {author} {\bibinfo {author} {\bibfnamefont {C.-H.}\ \bibnamefont
  {Du}}, \bibinfo {author} {\bibfnamefont {M.~E.}\ \bibnamefont {Ghazi}},
  \bibinfo {author} {\bibfnamefont {Y.}~\bibnamefont {Su}}, \bibinfo {author}
  {\bibfnamefont {I.}~\bibnamefont {Pape}}, \ and\ \bibinfo {author}
  {\bibfnamefont {P.~D.}\ \bibnamefont {Hatton}},\ }\href@noop {} {\bibfield
  {journal} {\bibinfo  {journal} {Phys. Rev. Lett.},\ }\textbf {\bibinfo
  {volume} {84}},\ \bibinfo {pages} {3911} (\bibinfo {year}
  {2000})}\BibitemShut {NoStop}%
\bibitem [{\citenamefont {Katsufuji}\ \emph {et~al.}(1996)\citenamefont
  {Katsufuji}, \citenamefont {Tanabe}, \citenamefont {Ishikawa}, \citenamefont
  {Fukuda}, \citenamefont {Arima},\ and\ \citenamefont {Tokura}}]{katsu;prb96}%
  \BibitemOpen
  \bibfield  {author} {\bibinfo {author} {\bibfnamefont {T.}~\bibnamefont
  {Katsufuji}}, \bibinfo {author} {\bibfnamefont {T.}~\bibnamefont {Tanabe}},
  \bibinfo {author} {\bibfnamefont {T.}~\bibnamefont {Ishikawa}}, \bibinfo
  {author} {\bibfnamefont {Y.}~\bibnamefont {Fukuda}}, \bibinfo {author}
  {\bibfnamefont {T.}~\bibnamefont {Arima}}, \ and\ \bibinfo {author}
  {\bibfnamefont {Y.}~\bibnamefont {Tokura}},\ }\href@noop {} {\bibfield
  {journal} {\bibinfo  {journal} {Phys. Rev. B},\ }\textbf {\bibinfo {volume}
  {54}},\ \bibinfo {pages} {R14230} (\bibinfo {year} {1996})}\BibitemShut
  {NoStop}%
\bibitem [{\citenamefont {Proffen}\ \emph {et~al.}(2002)\citenamefont
  {Proffen}, \citenamefont {Egami}, \citenamefont {Billinge}, \citenamefont
  {Cheetham}, \citenamefont {Louca},\ and\ \citenamefont
  {Parise}}]{proff;apa01i}%
  \BibitemOpen
  \bibfield  {author} {\bibinfo {author} {\bibfnamefont {T.}~\bibnamefont
  {Proffen}}, \bibinfo {author} {\bibfnamefont {T.}~\bibnamefont {Egami}},
  \bibinfo {author} {\bibfnamefont {S.~J.~L.}\ \bibnamefont {Billinge}},
  \bibinfo {author} {\bibfnamefont {A.~K.}\ \bibnamefont {Cheetham}}, \bibinfo
  {author} {\bibfnamefont {D.}~\bibnamefont {Louca}}, \ and\ \bibinfo {author}
  {\bibfnamefont {J.~B.}\ \bibnamefont {Parise}},\ }\href@noop {} {\bibfield
  {journal} {\bibinfo  {journal} {Appl. Phys. A},\ }\textbf {\bibinfo {volume}
  {74}},\ \bibinfo {pages} {s163} (\bibinfo {year} {2002})}\BibitemShut
  {NoStop}%
\bibitem [{\citenamefont {Peterson}\ \emph {et~al.}(2000)\citenamefont
  {Peterson}, \citenamefont {Gutmann}, \citenamefont {Proffen},\ and\
  \citenamefont {Billinge}}]{peter;jac00}%
  \BibitemOpen
  \bibfield  {author} {\bibinfo {author} {\bibfnamefont {P.~F.}\ \bibnamefont
  {Peterson}}, \bibinfo {author} {\bibfnamefont {M.}~\bibnamefont {Gutmann}},
  \bibinfo {author} {\bibfnamefont {T.}~\bibnamefont {Proffen}}, \ and\
  \bibinfo {author} {\bibfnamefont {S.~J.~L.}\ \bibnamefont {Billinge}},\
  }\href@noop {} {\bibfield  {journal} {\bibinfo  {journal} {J. Appl.
  Crystallogr.},\ }\textbf {\bibinfo {volume} {33}},\ \bibinfo {pages} {1192}
  (\bibinfo {year} {2000})}\BibitemShut {NoStop}%
\bibitem [{\citenamefont {Farrow}\ \emph {et~al.}(2007)\citenamefont {Farrow},
  \citenamefont {Juh\'as}, \citenamefont {Liu}, \citenamefont {Bryndin},
  \citenamefont {{Bo\v zin}}, \citenamefont {Bloch}, \citenamefont {Proffen},\
  and\ \citenamefont {Billinge}}]{farro;jpcm07}%
  \BibitemOpen
  \bibfield  {author} {\bibinfo {author} {\bibfnamefont {C.~L.}\ \bibnamefont
  {Farrow}}, \bibinfo {author} {\bibfnamefont {P.}~\bibnamefont {Juh\'as}},
  \bibinfo {author} {\bibfnamefont {J.}~\bibnamefont {Liu}}, \bibinfo {author}
  {\bibfnamefont {D.}~\bibnamefont {Bryndin}}, \bibinfo {author} {\bibfnamefont
  {E.~S.}\ \bibnamefont {{Bo\v zin}}}, \bibinfo {author} {\bibfnamefont
  {J.}~\bibnamefont {Bloch}}, \bibinfo {author} {\bibfnamefont
  {T.}~\bibnamefont {Proffen}}, \ and\ \bibinfo {author} {\bibfnamefont
  {S.~J.~L.}\ \bibnamefont {Billinge}},\ }\Doi {10.1088/0953-8984/19/33/335219}
  {\bibfield  {journal} {\bibinfo  {journal} {J. Phys: Condens. Mat.},\
  }\textbf {\bibinfo {volume} {19}},\ \bibinfo {pages} {335219} (\bibinfo
  {year} {2007})}\BibitemShut {NoStop}%
\bibitem [{\citenamefont {Larson}\ and\ \citenamefont {{Von
  Dreele}}(2004)}]{larso;unpub04}%
  \BibitemOpen
  \bibfield  {author} {\bibinfo {author} {\bibfnamefont {A.~C.}\ \bibnamefont
  {Larson}}\ and\ \bibinfo {author} {\bibfnamefont {R.~B.}\ \bibnamefont {{Von
  Dreele}}},\ }\href@noop {} {\enquote {\bibinfo {title} {General structure
  analysis system},}\ } (\bibinfo {year} {2004}),\ \bibinfo {note} {report No.
  LAUR-86-748, Los Alamos National Laboratory, Los Alamos, NM
  87545}\BibitemShut {NoStop}%
\bibitem [{\citenamefont {Toby}\ and\ \citenamefont
  {Billinge}(2004)}]{toby;aca04}%
  \BibitemOpen
  \bibfield  {author} {\bibinfo {author} {\bibfnamefont {B.~H.}\ \bibnamefont
  {Toby}}\ and\ \bibinfo {author} {\bibfnamefont {S.~J.~L.}\ \bibnamefont
  {Billinge}},\ }\href@noop {} {\bibfield  {journal} {\bibinfo  {journal} {Acta
  Crystallogr. A},\ }\textbf {\bibinfo {volume} {60}},\ \bibinfo {pages} {315}
  (\bibinfo {year} {2004})}\BibitemShut {NoStop}%
\bibitem [{\citenamefont {Wu}\ \emph {et~al.}(2002)\citenamefont {Wu},
  \citenamefont {Neumeier}, \citenamefont {Ling},\ and\ \citenamefont
  {Argyriou}}]{wu02}%
  \BibitemOpen
  \bibfield  {author} {\bibinfo {author} {\bibfnamefont {G.}~\bibnamefont
  {Wu}}, \bibinfo {author} {\bibfnamefont {J.~J.}\ \bibnamefont {Neumeier}},
  \bibinfo {author} {\bibfnamefont {C.~D.}\ \bibnamefont {Ling}}, \ and\
  \bibinfo {author} {\bibfnamefont {D.~N.}\ \bibnamefont {Argyriou}},\ }\Doi
  {10.1103/PhysRevB.65.174113} {\bibfield  {journal} {\bibinfo  {journal}
  {Phys. Rev. B},\ }\textbf {\bibinfo {volume} {65}},\ \bibinfo {pages}
  {174113} (\bibinfo {year} {2002})}\BibitemShut {NoStop}%
\bibitem [{\citenamefont {Tranquada}\ \emph {et~al.}(1995)\citenamefont
  {Tranquada}, \citenamefont {Lorenzo}, \citenamefont {Buttrey},\ and\
  \citenamefont {Sachan}}]{tran95b}%
  \BibitemOpen
  \bibfield  {author} {\bibinfo {author} {\bibfnamefont {J.~M.}\ \bibnamefont
  {Tranquada}}, \bibinfo {author} {\bibfnamefont {J.~E.}\ \bibnamefont
  {Lorenzo}}, \bibinfo {author} {\bibfnamefont {D.~J.}\ \bibnamefont
  {Buttrey}}, \ and\ \bibinfo {author} {\bibfnamefont {V.}~\bibnamefont
  {Sachan}},\ }\href@noop {} {\bibfield  {journal} {\bibinfo  {journal} {Phys.
  Rev. B},\ }\textbf {\bibinfo {volume} {52}},\ \bibinfo {pages} {3581}
  (\bibinfo {year} {1995})}\BibitemShut {NoStop}%
\bibitem [{\citenamefont {Guoqing}\ \emph {et~al.}(2002)\citenamefont
  {Guoqing}, \citenamefont {j.~j.}, \citenamefont {D.},\ and\ \citenamefont
  {N.}}]{Guoqi;prb02}%
  \BibitemOpen
  \bibfield  {author} {\bibinfo {author} {\bibfnamefont {W.}~\bibnamefont
  {Guoqing}}, \bibinfo {author} {\bibfnamefont {N.}~\bibnamefont {j.~j.}},
  \bibinfo {author} {\bibfnamefont {L.~C.}\ \bibnamefont {D.}}, \ and\ \bibinfo
  {author} {\bibfnamefont {A.~D.}\ \bibnamefont {N.}},\ }\Doi
  {10.1103/PhysRevB.65.174113} {\bibfield  {journal} {\bibinfo  {journal}
  {Phys. Rev. B},\ }\textbf {\bibinfo {volume} {65}},\ \bibinfo {pages}
  {174113} (\bibinfo {year} {2002})}\BibitemShut {NoStop}%
\bibitem [{\citenamefont {Hess}\ \emph {et~al.}(1999)\citenamefont {Hess},
  \citenamefont {B\"uchner}, \citenamefont {H\"ucker}, \citenamefont {Gross},\
  and\ \citenamefont {Cheong}}]{hess99}%
  \BibitemOpen
  \bibfield  {author} {\bibinfo {author} {\bibfnamefont {C.}~\bibnamefont
  {Hess}}, \bibinfo {author} {\bibfnamefont {B.}~\bibnamefont {B\"uchner}},
  \bibinfo {author} {\bibfnamefont {M.}~\bibnamefont {H\"ucker}}, \bibinfo
  {author} {\bibfnamefont {R.}~\bibnamefont {Gross}}, \ and\ \bibinfo {author}
  {\bibfnamefont {S.-W.}\ \bibnamefont {Cheong}},\ }\href@noop {} {\bibfield
  {journal} {\bibinfo  {journal} {Phys. Rev. B},\ }\textbf {\bibinfo {volume}
  {59}},\ \bibinfo {pages} {R10397} (\bibinfo {year} {1999})}\BibitemShut
  {NoStop}%
\bibitem [{\citenamefont {Billinge}\ \emph {et~al.}(1991)\citenamefont
  {Billinge}, \citenamefont {Davies}, \citenamefont {Egami},\ and\
  \citenamefont {Catlow}}]{billi;prb91}%
  \BibitemOpen
  \bibfield  {author} {\bibinfo {author} {\bibfnamefont {S.~J.~L.}\
  \bibnamefont {Billinge}}, \bibinfo {author} {\bibfnamefont {P.~K.}\
  \bibnamefont {Davies}}, \bibinfo {author} {\bibfnamefont {T.}~\bibnamefont
  {Egami}}, \ and\ \bibinfo {author} {\bibfnamefont {C.~R.~A.}\ \bibnamefont
  {Catlow}},\ }\href@noop {} {\bibfield  {journal} {\bibinfo  {journal} {Phys.
  Rev. B},\ }\textbf {\bibinfo {volume} {43}},\ \bibinfo {pages} {10340}
  (\bibinfo {year} {1991})}\BibitemShut {NoStop}%
\bibitem [{\citenamefont {Debye}(1912)}]{debye;adp12}%
  \BibitemOpen
  \bibfield  {author} {\bibinfo {author} {\bibfnamefont {P.}~\bibnamefont
  {Debye}},\ }\href@noop {} {\bibfield  {journal} {\bibinfo  {journal} {Ann.
  Phys.-Berlin},\ }\textbf {\bibinfo {volume} {39}},\ \bibinfo {pages} {789}
  (\bibinfo {year} {1912})}\BibitemShut {NoStop}%
\bibitem [{\citenamefont {Billinge}\ \emph {et~al.}(1996)\citenamefont
  {Billinge}, \citenamefont {DiFrancesco}, \citenamefont {Kwei}, \citenamefont
  {Neumeier},\ and\ \citenamefont {Thompson}}]{billi;prl96}%
  \BibitemOpen
  \bibfield  {author} {\bibinfo {author} {\bibfnamefont {S.~J.~L.}\
  \bibnamefont {Billinge}}, \bibinfo {author} {\bibfnamefont {R.~G.}\
  \bibnamefont {DiFrancesco}}, \bibinfo {author} {\bibfnamefont {G.~H.}\
  \bibnamefont {Kwei}}, \bibinfo {author} {\bibfnamefont {J.~J.}\ \bibnamefont
  {Neumeier}}, \ and\ \bibinfo {author} {\bibfnamefont {J.~D.}\ \bibnamefont
  {Thompson}},\ }\Doi {10.1103/PhysRevLett.77.715} {\bibfield  {journal}
  {\bibinfo  {journal} {Phys. Rev. Lett.},\ }\textbf {\bibinfo {volume} {77}},\
  \bibinfo {pages} {715} (\bibinfo {year} {1996})}\BibitemShut {NoStop}%
\bibitem [{\citenamefont {{Bo\v zin}}\ \emph {et~al.}(2010)\citenamefont {{Bo\v
  zin}}, \citenamefont {Malliakas}, \citenamefont {Souvatzis}, \citenamefont
  {Proffen}, \citenamefont {Spaldin}, \citenamefont {Kanatzidis},\ and\
  \citenamefont {Billinge}}]{bozin;s10}%
  \BibitemOpen
  \bibfield  {author} {\bibinfo {author} {\bibfnamefont {E.~S.}\ \bibnamefont
  {{Bo\v zin}}}, \bibinfo {author} {\bibfnamefont {C.~D.}\ \bibnamefont
  {Malliakas}}, \bibinfo {author} {\bibfnamefont {P.}~\bibnamefont
  {Souvatzis}}, \bibinfo {author} {\bibfnamefont {T.}~\bibnamefont {Proffen}},
  \bibinfo {author} {\bibfnamefont {N.~A.}\ \bibnamefont {Spaldin}}, \bibinfo
  {author} {\bibfnamefont {M.~G.}\ \bibnamefont {Kanatzidis}}, \ and\ \bibinfo
  {author} {\bibfnamefont {S.~J.~L.}\ \bibnamefont {Billinge}},\ }\href@noop {}
  {\bibfield  {journal} {\bibinfo  {journal} {Science},\ }\textbf {\bibinfo
  {volume} {330}},\ \bibinfo {pages} {1660} (\bibinfo {year}
  {2010})}\BibitemShut {NoStop}%
\bibitem [{\citenamefont {Lawler}\ \emph {et~al.}(2010)\citenamefont {Lawler},
  \citenamefont {Fujita}, \citenamefont {Lee}, \citenamefont {Schmidt},
  \citenamefont {Kohsaka}, \citenamefont {Kim}, \citenamefont {Eisaki},
  \citenamefont {Uchida}, \citenamefont {Davis}, \citenamefont {Sethna},\ and\
  \citenamefont {Kim}}]{lawl10}%
  \BibitemOpen
  \bibfield  {author} {\bibinfo {author} {\bibfnamefont {M.~J.}\ \bibnamefont
  {Lawler}}, \bibinfo {author} {\bibfnamefont {K.}~\bibnamefont {Fujita}},
  \bibinfo {author} {\bibfnamefont {J.}~\bibnamefont {Lee}}, \bibinfo {author}
  {\bibfnamefont {A.~R.}\ \bibnamefont {Schmidt}}, \bibinfo {author}
  {\bibfnamefont {Y.}~\bibnamefont {Kohsaka}}, \bibinfo {author} {\bibfnamefont
  {C.~K.}\ \bibnamefont {Kim}}, \bibinfo {author} {\bibfnamefont
  {H.}~\bibnamefont {Eisaki}}, \bibinfo {author} {\bibfnamefont
  {S.}~\bibnamefont {Uchida}}, \bibinfo {author} {\bibfnamefont {J.~C.}\
  \bibnamefont {Davis}}, \bibinfo {author} {\bibfnamefont {J.~P.}\ \bibnamefont
  {Sethna}}, \ and\ \bibinfo {author} {\bibfnamefont {E.-A.}\ \bibnamefont
  {Kim}},\ }\href@noop {} {\bibfield  {journal} {\bibinfo  {journal} {Nature},\
  }\textbf {\bibinfo {volume} {466}},\ \bibinfo {pages} {347} (\bibinfo {year}
  {2010})}\BibitemShut {NoStop}%
\end{thebibliography}

%

\end{document}

%\preprint{APS/123-QED}

%================================================================================================
\title{Evidence for Fluctuating Charge Stripes far above the Charge-Ordering Transition  in La$_{1.67}$Sr$_{0.33}$NiO$_{4}$ - Supplementary Material}
%================================================================================================
\author{A. M. Milinda Abeykoon}
\affiliation{Condensed Matter Physics and Materials Science Department, Brookhaven National Laboratory}

\author{Emil~S. Bo\v{z}in}
\affiliation{Condensed Matter Physics and Materials Science Department, Brookhaven National Laboratory}

\author{Wei-Guo Yin}
\affiliation{Condensed Matter Physics and Materials Science Department, Brookhaven National Laboratory}

\author{Genda Gu}
\affiliation{Condensed Matter Physics and Materials Science Department, Brookhaven National Laboratory}

\author{John Tranquada}
\affiliation{Condensed Matter Physics and Materials Science Department, Brookhaven National Laboratory}

\author{John P. Hill}
\affiliation{Condensed Matter Physics and Materials Science Department, Brookhaven National Laboratory}

\author{Simon J. L. Billinge}
\affiliation{Condensed Matter Physics and Materials Science Department, Brookhaven National Laboratory}
\affiliation{Department of Applied Physics and Applied Mathematics, Columbia University}
%%\author{N. E. others?}
%\author{}
%\affiliation{}
%\author{}
%\affiliation{}
%\author{}
%\affiliation{}

%------------------------------------------------------------------------------------------------------------------------

\begin{abstract}

\end{abstract}

%------------------------------------------------------------------------------------------------------------------------

\date{\today}% It is always \today, today,

%\pacs{, }% PACS, the Physics and Astronomy

\maketitle

\section{Sample preparation}
The powder sample of La$_{1.67}$Sr$_{0.33}$NiO$_{x}$ used in this
study was initially prepared as a single crystal by the
traveling-solvent floating-zone method using high purity starting materials at Brookhaven National Laboratory. La$_{2}$O$_{3}$, SrCO$_{3}$ and NiO powders of 99.999\% purity were mixed in appropriate molar ratios in an agate mortar, and the mixture was initially calcined for 24 hours at 1200 $^{\circ}$C. The calcined material was then reground and further calcined for additional 48 hours at 1300 $^{\circ}$C for feed rods. The product was pulverized again, placed into a rubber tube and hydrostatically pressed under 4200 Kg/cm$^2$ load to form a rod. Such pressed rod was sintered in the air for additional 72 hours at 1450 $^{\circ}$C, and slow cooled down in the furnace. Finally, sintered rod was finely pulverized to be used in neutron diffraction experiments.

\section{Neutron diffraction measurements}
Neutron time-of-flight diffraction measurements were carried out at the NPDF beam-line~\cite{proff;apa01i} at Los Alamos Neutron Science Center (LANSCE) at Los Alamos National Laboratory. Finely pulverized sample (approximate mass of 15~grams) was loaded into an extruded cylindrical vanadium container in He atmosphere which serves as an exchange gas, and sealed. Sample temperature was controlled using a He closed-cycle cryo-furnace, which has an operating range from 15~K to 500~K. Neutron diffraction data were collected at 22 different temperatures for 3~hours at each temperature, with 15 minute equilibration time between the successive measurements. Standard data reduction to obtain the pair distribution function (PDF) was carried out using the program PDFgetN~\cite{peter;jac00}.

\section{Data analysis}
\subsection{Structural modeling}
Sequential Rietveld refinements~\cite{rietv;ac67, young;b;trm93, mccus;jac99, vondr;jac07} were carried out %and a sequential PDF analysis were carried out on
on LSNO neutron powder diffraction data using the GSAS~\cite{larso;unpub87} software operated by EXPGUI\cite{toby;aca04} platform %and PDFGui
to explore SRO or LRO structural responses to charge order in the system. Tetragonal symmetry (space group I4/mmm) with initial lattice parameters a~=~b~=~3.82~\AA\ and c~=~12.72~\AA\ was used as the LSNO starting model. Unit cell parameters, fractional coordinates, anisotropic atomic displacement parameters, polynomial background function and the overall scale parameter were refined sequentially to obtain satisfactory agreement with the data. Sample absorption and adequate resolution function parameters (TOF profile function 1) were refined against the 10~K dataset, and then kept fixed at such refined values throughout the sequential refinement of the temperature dependent data. Complementary sequential analysis on the PDFs derived from the same data was also performed over variable lengthscales using the software package PDFgui~\cite{farro;jpcm07}, yielding results consistent with the Rietveld analysis. In addition, model independent PDF peak height analysis of selected PDF features was performed to verify the observed responses to CO in the temperature dependencies of ADPs obtained in structural modeling.
\subsection{The Debye model of ADP temperature dependence}
Temperature evolution of ADPs in materials can typically be well described using the Debye model~\cite{debye;adp12}, given by
%=========================
 \begin{equation}
%\begin {math}
U^{2}(T)=\frac { 3h^2} {4\pi^{2}mk_{B}\Theta_{D}}\left(\frac{\Phi(\Theta_{D}/T)}{\Theta_{D}/T} +\frac{1}{4}\right)+U^{2}_{0},
\label{eq:DebyeModel1}
%\end{math}
\end{equation}
%=========================
where h is the Plank constant, m is the atomic mass, k$_{B}$  is the Boltzman
constant, $\Theta_{D}$ is the Debye temperature, and 
%=========================
 \begin{equation}
%\begin {math}
\Phi(x)=\frac{1}{x} \int^{x}_{0} \frac{x'dx'}{e^{x'}-1}.
\label{eq:DebyeModel2}
%\end{math}
\end{equation}
%=========================
%
Qualitative interrogation of how well the Debye model fits the temperature dependencies of the ADPs obtained from structural refinements can be rather informative. This is particularly useful in sensing the development of broken symmetry states that are lacking long range ordering and whose structural response is escaping detection using conventional crystallographic approaches. Any deviation from the canonical behavior described by the model, when observed, is typically ascribed either to the overall inadequacy of the structural model used, or to the presence of nanoscale features that deviate from the average structure. In the particular case of LSNO, CO related SRO structural modulations which are not captured in the average structural model reveal themselves as deviations of the refined ADPs from the Debye-type behavior, as discussed in the Main Article.
\subsection{Estimate of the effective pseudogap}
\begin{figure}[h!]
\center
\includegraphics[scale=0.54]{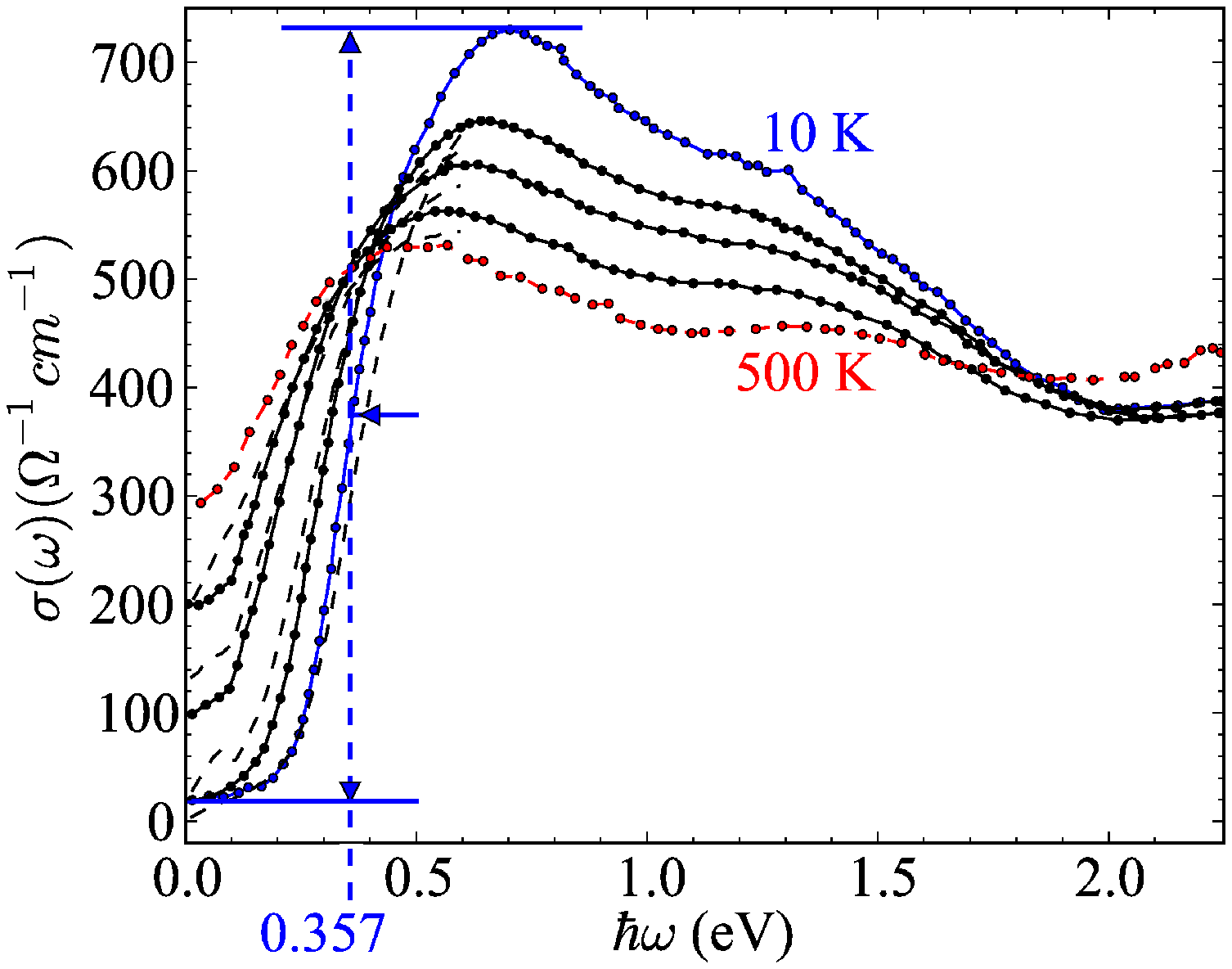}
\caption {(Color online) Illustration of protocol used to estimate the effective pseudogap of LSNO using digitized optical conductivity data from Katsufuji et. al~\cite{katsu;prb96}. Data span temperature range from 10~K (blue) to 500~K (red).}
\label{fig:LSNO_OPC}
\end{figure}

As discussed in the Main Article, we have evaluated the energy at which the optical conductivity falls to half of its peak value at each temperature. Figure~\ref{fig:LSNO_OPC} explains the protocol used to achieve this. Optical conductivity data for LSNO was first digitized from the original work of Katsufuji et. al~\cite{katsu;prb96}.
%=====================

%=====================
Then for such data, as illustrated for the case of 10~K, the minimum and maximum values of optical conductivity are identified (solid horizontal lines), and the value of the energy corresponding to half the value of the peak-to-base optical conductivity is recorded for all available temperatures. This was then normalized by the value obtained for the base temperature, and plotted against the SRO parameter and shown in the Main Article.

\section{Lattice energy calculated from electrostatic considerations with and without ordered charge stripes}
In LSNO the $i$th-neighbor interlayer Ni-Ni distance, $r_{i}$, is given by
\begin{equation}
\label{r}
  r_i = \sqrt{c^2/4+(n/2)a^2} \\
\end{equation}
where $a=3.8268$ {\AA} (the nearest Ni-Ni distance) and $c=12.7086$ {\AA} are the lattice constants and $n=1$, 5, 9 and 13 for $r_1$, $r_2$, $r_3$ and $r_4$, respectively.  Finally, $r_5=c$.  %The $c$/$a$ ratio is approximately 3.
 
\begin{figure}[tb]
\center
\includegraphics[width=1\columnwidth,clip=true,angle=0]{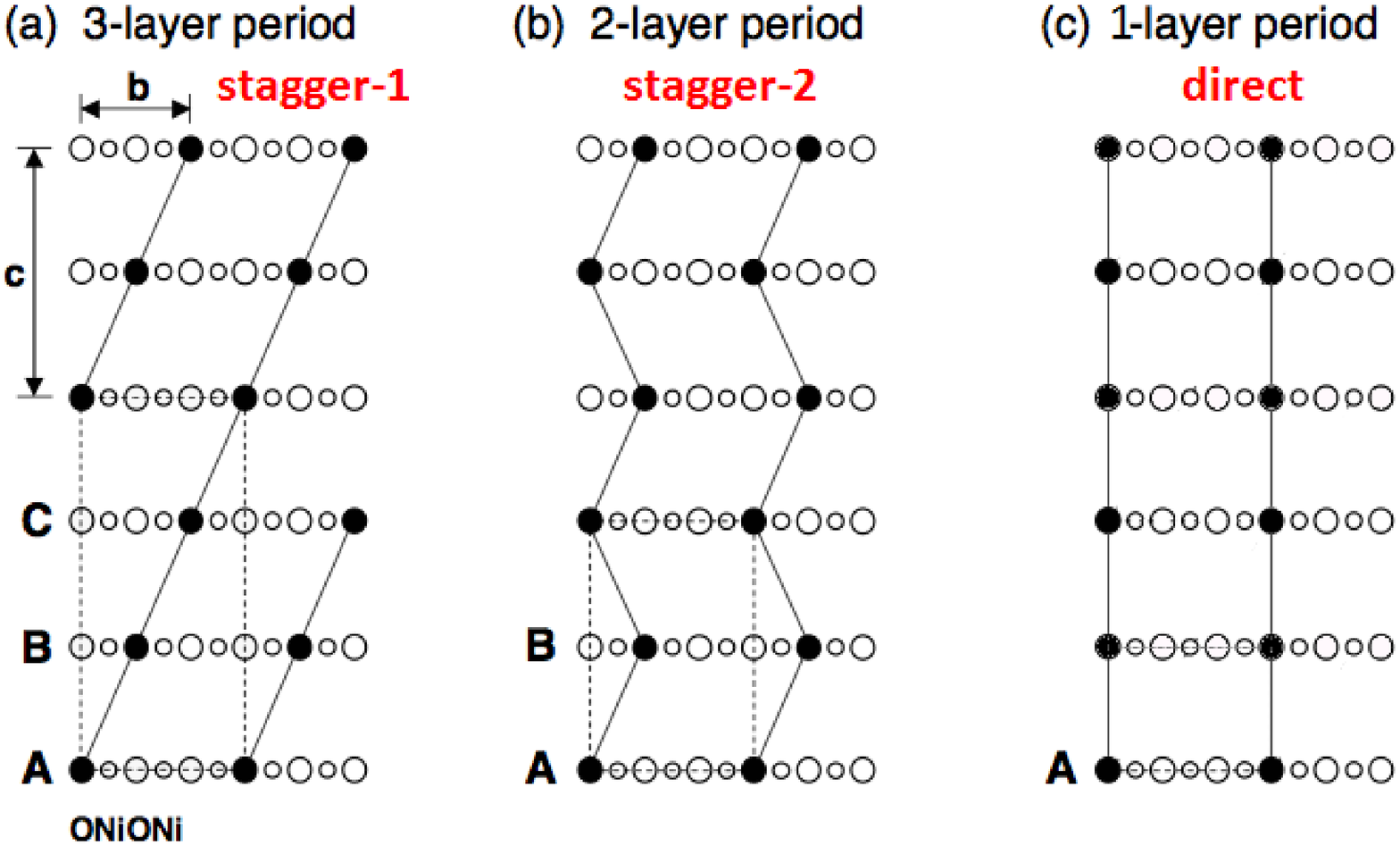}
\caption{\label{fig:stack}Stripe stacking patterns. (a) three-layer period comparable to cubic stacking, (b) two-layer period comparable to hexagonal stacking, and (c) one-layer period.}
\end{figure}

The observed charge stripes in the low-$T$ phase of LSNO for $x=1/3$ lie in the NiO$_2$ plane and are diagonal with respect to the nearest neighbor nickel ions. They are not directly stacked along the $c$ axis~\cite{hucke;prb06} but are staggered.  There are two possible ways to stagger them, as illustrated in Fig.~\ref{fig:stack}(a) and (b).
%========================
%========================
In LSNO these two patterns seem to be degenerate, as they appear randomly~\cite{hucke;prb06}. This suggests that the screened interlayer Coulomb interaction, $V(r)$, beyond the distance $c$ is negligible. That is, it is sufficient to consider $V(r)$ up to $r_4$. These two patterns will be referred to as ``stagger'', and the one with directly stacked charge stripes along the $c$ axis (Fig.~\ref{fig:stack}(c)) will be referred to as ``direct''. The energies per Ni contributed by $V(r)$ for the ``stagger'' and ``direct'' patterns are
\begin{eqnarray}
E_\mathrm{stagger}&=&2[V(r_1)+4V(r_2)+2V(r_4)] \nonumber \\
E_\mathrm{direct}&=&2[2V(r_1)+4V(r_3)+4V(r_4)].
\end{eqnarray}
It follows that
\begin{equation}
E_\mathrm{stagger}-E_\mathrm{direct}=2[-V(r_1)+4V(r_2)-4V(r_3)-2V(r_4)]<0.
\end{equation}
 
For the high-$T$ normal state, where charges are distributed on Ni uniformly, the energy per Ni contributed by $V(r)$ is
\begin{eqnarray}
E_\mathrm{normal}&=& \frac{2}{x} x^2 [4V(r_1)+8V(r_2)+4V(r_3)+8V(r_4)] \nonumber\\
&=& \frac{2}{3}[4V(r_1)+8V(r_2)+4V(r_3)+8V(r_4)].
\end{eqnarray}
Then, we arrive at
\begin{eqnarray}
E_\mathrm{stagger}-E_\mathrm{normal}&=&\frac{2}{3}[-V(r_1)+4V(r_2)-4V(r_3)-2V(r_4)] \nonumber \\
&=& \frac{1}{3}(E_\mathrm{stagger}-E_\mathrm{direct}) < 0.
\end{eqnarray}
 
For the intermediate-$T$ regime, where charge stripes survive but loose stacking order along the $c$ axis, the energy per Ni contributed by $V(r)$ is
\begin{equation}
E_\mathrm{fluc}=\frac{2}{3}E_\mathrm{stagger}+\frac{1}{3}E_\mathrm{direct}.
\end{equation}
Then, we get
\begin{eqnarray}
E_\mathrm{stagger}&-&E_\mathrm{fluc}=\frac{1}{3}(E_\mathrm{stagger}-E_\mathrm{direct}) < 0, \\
E_\mathrm{fluc}&-&E_\mathrm{normal}= 0.
\end{eqnarray}
These relations are exact regardless of the actual form of $V(r)$. They mean that as $T$ drops, the formation of the static charge stripe order will noticeably lower the energy contributed by the interlayer electrostatic interaction, but the formation of the fluctuating charge stripes will not. Hence, we anticipate an anomalous reduction of $c/a$ at $T_\mathrm{CO}$, but not at the crossover between the intermediate- and high-$T$ regimes.

\section {Acknowledgements}
Work at Brookhaven is supported by the Office of Basic Energy Sciences, Division of Materials Sciences and Engineering, U.S. Department of Energy through account DE-AC02-98CH10886. This work has benefited from the use of NPDF at LANSCE, funded by DOE Office of Basic Energy Sciences. LANL is operated by Los Alamos National Security LLC under DOE Contract No. DE-AC52-06NA25396.

%\bibliography{billinge-group,abb-billinge-group,everyone}

%